\documentstyle[emlines,11pt]{article}
\title{PRINCIPLES OF A UNIFIED THEORY OF SPACETIME AND PHYSICAL
INTERACTIONS}
\author{Yu.S. Vladimirov \\
Physics Faculty, Moscow State University}
\date{}
\begin{document}
\maketitle
\begin{abstract}
Principles of a new approach (binary geometrophysics) are presented to
construct the unified theory of spacetime and the familiar kinds of physical
interactions. Physically, the approach is a modified S-matrix theory
involving ideas of the multidimensional geometric models of physical
interactions of Kaluza-Klein's type as well as Fokker-Feynman's
action-at-a-distance theory. Mathematically, this is a peculiar binary
geometry being described in algebraic terms. In the present approach the
binary geometry volume is a prototype of three related notions: the S-matrix,
the physical action (Lagrangian) of both strong and electroweak interactions,
and the multidimensional metric. A transition from microworld geometrophysics
to the conventional physical theory in classical spacetime are characterized.
\end{abstract}

\section{Introduction}
\hspace*{\parindent}
In attempts to construct microworld physics for the last half of a century
hopes have been pinned successively on a few key ideas and principles.
So, in the fifties those were the principles of quantum field theory, in the
sixties attempts were made to construct the S-matrix approach and quantum
theory axiomatics, in the seventies and early eighties a primary emphasis was
placed upon group methods and gauge theory of physical interactions, at the
very end of the 20th century hopes were pinned first on supesymmetric
theories and then on superstrings and p-branes. At the same time,
other (side) concepts: Fokker-Feynman's
action-at-a-distance theory, Penrose's twistor theory, multidimensional
geometrical models of Kaluza-Klein's type etc. were investigated.

In all these investigations a relation to the coordinate spacetime was
formulated in any event. The latter was a priori assumed four-dimensional,
either was generalized to superspaces or multidimensional manifolds, or it
was considered that for constructing microworld physics needless to proceed
from the coordinate spacetime, and the momentum space should be used as the
basis for constructing a theory (as was declared by Chew in the S-matrix
approach [1]), either it was proposed to start from complex notions of
another kind, e.g. from twistors in Penrose's program [2, 3], then arriving
at both quantum theory and the classical spacetime.

In the present paper a new approach is proposed to construct microworld
physics, involving a number of ideas and methods of the previous research and
allowing one to consider the essence of spacetime and physical interactions
from another viewpoint. The theory being put forward is called {\it binary
geometrophysics}. Its premises have somewhat in common with Penrose's twistor
program, however our starting points are more abstract and richer in their
consequences. Binary geometrophysics is closest to the S-matrix theory. The
word ``binary'' reflects two sets of states of the system, i.e. the initial
(i) and the final (out) ones, forming the basis of the theory. The theory
itself is a peculiar binary geometry describing microworld physics.

As known, the idea of S-matrix approach was put forward by J. Wheeler [4] and
W. Heisenberg [5] and was developed into the S-matrix in the sixties in the
works by a number of authors (e.g., see [1, 6]). Its basis constituted
principles of Lorentz invariance, analyticity, causality etc. Particles and
their characteristics were related to poles in the complex plane.

Binary geometrophysics and the S-matrix theory share, apart from the two
kinds of states, a possibility of constructing a theory ``independently of
whether microscopic spatial-temporal continuum does exist or not'' [1, p.17].
In binary geometrophysics there is no a priori spacetime for microworld. It
arises only in the relations between macro-objects (according to the idea of a
macroscopic nature of the classical spascetime [7]). On the very elementary
level there becomes invalid the differential and integral calculus alongside
spatial-temporal continuum. There remain only algebraic methods.

To construct binary geometrophysics, the mathematical (algebraic) formalism
of Yu.I. Kulakov's binary physical structures [8, 9, 10] has been used. (The
latter was previously developed for other purposes.) As in the S-matrix
theory, of importance here is a symmetry, however, in our theory there
emerges a more general than the Lorentzian symmetry from which one may turn
to Lie groups, in particular, to the Lorentz group and internal symmetries
being used in gauge theories.

In the present approach the function continuity condition is used only at the
first stage to find algebraic laws of structures (relation systems), and then
from merely algebraic considerations one may arrive at a definition of
particles and a relationship between constants and charges in strong and
electroweak intractions. But above all, in the framework of binary
geometrophysics, from the unified expressions prototypes of three
related notions: the S-matrix, the action (Lagrangian) for strong and
electroweak interactions of elementary particles and the multidimensional
metric are derived.

The ideas and methods used in the multidimensional geometrical models of
physical interactions of Kaluza-Klein's type theories (e.g., see [13, 14])
and Fokker-Feynman's action-at-a-distance theory [15--17] prove necessary in
constructing binary geometrophysics [11, 12].

It should be emphasized that binary geometrophysics has a relational pattern.
Relations between elements, i.e. primary notions of the theory, play the key
part in it. Specifically, the microanalogue of reference frames in Relativity
[18], called a binary system of complex relations (BSCR), is put in the
forefront. If micro-oobjects are denoted by the symbol $\mu$, and
macro-objects by the symbol $m$, then the starting points of the
{\it relational} theory should be defined by the symbol $R_{\mu}(\mu)$, where
$\mu$ in parentheses means that micro-objects are considered in it, and the
subscript $\mu$ means that this is done with respect to micro-objects as well.
In such terms classical mechanics, describing macro-objects with respect to
macroinstruments, is denoted as $R_m(m)$, and quantum mechanics -- as
$R_m(\mu)$.

The first two sections present principal ideas and principles of binary
geometrophysics at the level $R_{\mu}(\mu)$ in the absence of notions of the
classical spacetime, and in the fourth section a transition to the
conventional theory $R_m(m)$ in the presence of these notions is discussed.

\section{Binary Geometry of the Microworld}
\hspace*{\parindent}
Briefly outline the principal notions of BSCR forming the basis of the theory
$R_{\mu}(\mu)$ and give their physical interpretation.

{\bf 1. Two Sets of Elements}

It is postulated that there are two sets of elements. Denote the first set by
a symbol ${\cal M}$, and the second one by ${\cal N}$. Elements of the first
set are denoted by Latin letters ($i,j,k,\ldots$), and the elements of the
second one -- by Greek letters ($\alpha, \beta, \gamma,\ldots$). Between any
pair of the elements of different sets a pair relation -- some complex number
$u_{i\alpha}$ is given (see Fig. 1). The elements of the two sets have the
following physical meaning. The elements of the first set ${\cal M}$
characterize {\it initial states} of particles, and the elements of the
second one ${\cal N}$ -- {\it final states}. Thus, an idea of the S-matrix
approach (more exactly, an elementary chain of any evolution), i.e an
initial-to-final transition, proves to be laid in the very fundamental
notions of BSCR. The complex relations between elements in the two states are
a prototype of both the amplitudes of a transition between states and the
notions of momenta, and finally particle coordinates.

{\bf 2. Law and Fundamental Symmetry}

It is postulated that there exists some algebraic {\it law}, connecting all
possible relations between any $r$ elements of the set ${\cal M}$ and $s$
elements of the set ${\cal N}$:
\begin{equation} \Phi_{(r,s)}(u_{i\alpha}, u_{i\beta},\ldots,
u_{k\gamma})=0. \end{equation}

The integers $r$ and $s$ characterizes {\it rank} ($r,s$) of BSCR. The
essential point of the theory is a requirement of {\it fundamental symmetry}
lying in the law (1) to be valid, while substituting the given set of
elements by any others in the corresponding sets. The fundamental symmetry
and the continuity condition allow functional-differential equations to be
derived from them and the form of both pair relations $u_{i\alpha}$ as well
as the function $\Phi_{(r,s)}$ itself to be found (see [8, 9, 10]).
\begin{figure}[htbp]
\special{em:linewidth 0.4pt}
\unitlength 1mm
\linethickness{0.4pt}
\begin{center}
\begin{picture}(92.33,55.00)
\put(55.17,15.50){\oval(69.67,19.00)[]}
\put(55.00,45.00){\oval(70.00,20.00)[]}
\emline{29.33}{18.00}{1}{29.33}{44.00}{2}
\emline{29.33}{44.00}{3}{39.67}{18.33}{4}
\emline{39.67}{18.33}{5}{39.67}{43.67}{6}
\emline{39.67}{43.67}{7}{51.33}{18.00}{8}
\emline{51.33}{18.00}{9}{51.67}{43.67}{10}
\emline{51.67}{43.67}{11}{29.33}{17.67}{12}
\emline{29.33}{17.67}{13}{39.67}{44.00}{14}
\emline{64.00}{17.67}{15}{64.00}{43.67}{16}
\emline{64.00}{43.67}{17}{75.33}{17.67}{18}
\emline{75.33}{17.67}{19}{75.33}{43.67}{20}
\emline{75.33}{43.67}{21}{64.00}{18.00}{22}
\emline{64.00}{18.00}{23}{29.33}{44.33}{24}
\emline{29.33}{44.33}{25}{75.33}{18.00}{26}
\emline{75.33}{43.33}{27}{29.33}{17.00}{28}
\emline{29.33}{17.00}{29}{64.00}{43.33}{30}
\emline{51.67}{43.33}{31}{39.67}{18.33}{32}
\put(69.33,30.67){\oval(20.67,33.33)[]}
\put(29.33,17.33){\circle*{2.00}}
\put(40.00,17.67){\circle*{2.00}}
\put(51.33,17.67){\circle*{2.00}}
\put(64.00,17.67){\circle*{2.00}}
\put(75.33,18.00){\circle*{2.00}}
\put(85.00,17.67){\circle*{2.00}}
\put(29.33,44.00){\circle*{2.00}}
\put(39.67,43.67){\circle*{2.00}}
\put(51.67,43.67){\circle*{2.00}}
\put(64.00,43.67){\circle*{2.00}}
\put(75.33,43.33){\circle*{2.00}}
\put(85.33,43.67){\circle*{2.00}}
\put(24.67,12.67){\dashbox{3.00}(30.67,9.33)[cc]{}}
\put(24.67,39.00){\dashbox{3.00}(30.67,8.67)[cc]{}}
\put(14.00,15.33){\makebox(0,0)[cc]{${\cal M}$}}
\put(14.00,44.33){\makebox(0,0)[cc]{${\cal N}$}}
\put(39.67,51.00){\makebox(0,0)[cc]{s elements}}
\put(39.67,9.33){\makebox(0,0)[cc]{r elements}}
\put(69.33,9.67){\makebox(0,0)[cc]{BSCR basis}}
\put(32.00,14.67){\makebox(0,0)[cc]{$i$}}
\put(42.67,15.00){\makebox(0,0)[cc]{$k$}}
\put(53.67,15.00){\makebox(0,0)[cc]{$j$}}
\put(66.33,15.33){\makebox(0,0)[cc]{$m$}}
\put(72.33,15.67){\makebox(0,0)[cc]{$n$}}
\put(82.33,15.67){\makebox(0,0)[cc]{$l$}}
\put(32.33,45.67){\makebox(0,0)[cc]{$\alpha$}}
\put(42.67,45.33){\makebox(0,0)[cc]{$\beta$}}
\put(54.00,46.00){\makebox(0,0)[cc]{$\gamma$}}
\put(66.33,45.00){\makebox(0,0)[cc]{$\mu$}}
\put(72.67,45.00){\makebox(0,0)[cc]{$\nu$}}
\put(82.33,45.33){\makebox(0,0)[cc]{$\lambda$}}
\put(26.33,30.67){\makebox(0,0)[cc]{$u_{i\alpha}$}}
\put(66.00,31.00){\makebox(0,0)[cc]{$u_{m\mu}$}}
\end{picture}
\end{center}
\caption{Binary system of relations (structure) of rank $(r,s)$.}
\end{figure}
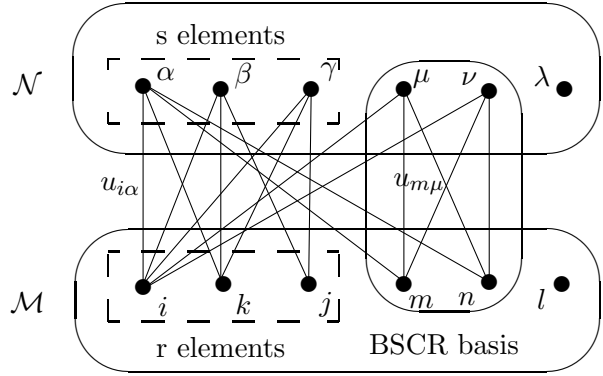

In binary geometrophysics BSCR of symmetric ranks $(r,r)$ are used, with
nondegenerate and degenerate systems of relations being distinguished. For
the nongeranerate BSCR the law is written via a determinant of pair
relations:
\begin{equation}
\Phi_{(r,r)}(u_{i\alpha},u_{i\beta},\ldots)=
\left|\begin{array}{cccc}
u_{i\alpha} & u_{i\beta} & \cdots & u_{i\gamma} \\
u_{k\alpha} & u_{k\beta} & \cdots & u_{k\gamma} \\
\cdots & \cdots & \cdots & \cdots \\
u_{j\alpha} & u_{j\beta} & \cdots & u_{j\gamma}
\end{array}\right|=0, \end{equation}
where the pair relations are represented in the form
\begin{equation} u_{i\alpha}=\sum_{l=1}^{r-1}i^l\alpha^l.
\end{equation}
Here $i^1, \ldots, i^{r-1}$ -- $(r-1)$ parameters of the element
$i$, and $\alpha^1,\ldots$, $\alpha^{r-1}$ -- $(r-1)$ parameters of the
element $\alpha$.

{\bf 3. Elementary Basis}

The origin of the element parameters should be especially dwelled on. They
are analogues of the notions of coordinates in the conventional
geomentry. To arrive at them, in the law (1) one should separate $r-1$
elements of the set ${\cal M}$ and $s-1$ elements of the set ${\cal N}$ and
consider them to be {\it standard}. Fig. 1 denotes these elements by letters
$m$, $n$, $\mu$, $\nu$. Then this law may be interpreted as a relationship
determining a pair relation between two nonstandard elements (say, the
elements $i$ and $\alpha$) in terms of their relations to standard elements
(see Fig. 1). Relations between the standard elements themselves may be
considered given forever. Then the pair relation $u_{i\alpha}$ proves to be
characterized by $s-1$ parameters of the element $i$ (its relations to $s-1$
standard elements of the set ${\cal N}$) and similar $r-1$ parameters of the
element $\alpha$. The system of the standard elements comprises
{\it elementary basis of BSCR}.

{\bf 4. Fundamental Relations}

In the BSCR of rank $(r,r)$ an important part is played by {\it fundamental}
$(r-1)\times (r-1)$-{\it relations}, being nonzero minors of the maximal
order\footnote[1]{It should be especially noted that this theory does not
introduce anything from outside but uses only those notions which naturally
arise in the framework of different rank BSCR.} in the determinant of the law
(2). For BSCR of any rank $(r,r)$ the fundamenta relations are written in
terms of a product of two determinants composed of parameters of one sort:
\begin{equation} \left[{\alpha\atop i}{\beta\atop
k}{\cdots\atop\cdots}\right]
\equiv \left|\begin{array}{ccc} u_{i\alpha} & u_{i\beta} &
\cdots \\ u_{k\alpha} & u_{k\beta} & \cdots \\ \cdots & \cdots
& \cdots \end{array}\right|=
\left|\begin{array}{ccc} i^1 & k^1 & \cdots \\ i^2 & k^2 &
\cdots \\ \cdots & \cdots & \cdots \end{array}\right|\times
\left|\begin{array}{ccc}
\alpha^1 & \beta^1 & \cdots \\ \alpha^2 & \beta^2 & \cdots \\
\cdots & \cdots & \cdots \end{array}\right|.
\end{equation}

{\bf 5. Linear Transformations}

Transitions from one elementary basis to another may be shown (see [11])
to be described by linear transformations of element parameters of two sets:
\begin{equation} i^{\prime s}= C^s_r i^r; \ \ \ \alpha^{\prime
s}=C^{\star s}_r \alpha^r, \end{equation}
where $C^s_r$ и $C^{\star s}_r$ are the coefficients defining a class of
binary systems (standard elements) being used. Restrict ourselves to the case
of complex conjugate coefficients.

{\bf 6. Two-Component and Finsler Spinors}

Low-rank BSCR notions are, in fact, used in modern theoretical physics. In
In particular, the two-component spinor theory [19] naturally arises in the
framework of the minimamal nondegenerate rank (3,3) BSCR. Really, according
to this theory, the elements are characterized by pairs of complex parameters
$ i\rightarrow (i^1, i^2)$, $ \alpha\rightarrow
(\alpha^1, \alpha^2)$, i.e. are vectors of the two-dimensional complex space
where the linear transformations (5) are defined. Restrict ourselves to the
liner transformations leaving invariant each of the $2\times 2$ determinants
on the left in (4). Since each of such determinants is an antisymmetric
bilinear form: $ i^1 k^2-i^2 k^1=Inv; \alpha^1\beta^2-\alpha^2\beta^1=Inv$,
the definition of two-component spinors becomes evident. For a selected
class of transformations (5) the coefficients satisfy the condition
$C^1_1C^2_2-C^1_2C^2_1=1,$ i.e. these transformations belong to a
six-parameter group $SL(2,C)$. Such transformations single out the privileged
class of standard elements corresponding to inertial frames of reference in
General Relativity.

In the framework of rank (3,3) BSCR, the transformations keeping invariant
both the antisymmetric forms on the right in (4) and the pair relations
$u_{i\alpha}$ form the three-parameter group $SU(2)$ corresponding to
rotations in three-dimensional space. As follows from the foregoing, one may
state that {\it four-dimensionality of the classical spacetime and the
signature} $(+ - - -)$ {\it are due to the rank 3,3) of the first
nondegenerate BSCR}.

At this level of theory development in the framework of rank (3,3) BSCR we,
in fact, arrive at the notions comprising Penrose's twistor theory. However
in binary geometrophysics, relation systems of a higher rank $(r,r)$ are
considered. In them the elements are described by $r-1$-dimensional vectors.
Requiring the corresponding antisymmetric forms in (4) under the
transformations (5), we arrive at a nonstandard generalization\footnote[2]
{It is conventional to use a generalization of two-component spinors based on
Clifford's algebras over the field of real numbers (e.g.,see [20]), where the
spinors have $2^n$ components.} of the two-component spinors which are
naturally called {\it Finsler $(r-1)$-component spinors} [21]. These
transformations form a group $SL(r-1,C)$.

{\bf 7. Description of Elementary Particles}

In binary geometrophysics the elementary particles are described by a few
elements in each of the two sets. In the framework of a simplified model
based on the rank (3,3) BSCR, massive leptons (electrons) are described by
pairs of elements, and neutrinos -- by one element in each of the sets.
Let the electron $(e)$ be described by two pairs of elements: $i$, $k$ and
$\alpha$, $\beta$, then it may be characterized in terms close to the
conventional ones, i.e. by the four-component column and line:
\begin{equation} e=\left(\begin{array}{c}
i^1 \\ i^2 \\ \beta_1 \\ \beta_2 \end{array}\right)=
\left(\begin{array}{c}
i^1 \\ i^2 \\ \beta^2 \\ -\beta^1 \end{array}\right); \ \ \
e^{\dag}=(\alpha^1, \alpha^2, k_1, k_2), \end{equation}
where the quantities with subscripts denote covariant components of
spinors.

On the elements describing the leptons imposed are constraints along the
horizontal (in one set of elements) and along the vertical (in two sets).
Thus, for {\it free} leptons the vertical constraints mean that the
parameters of two pairs of elements in the two sets are complex conjugate to
one another, following the rules of quantum mechanics. For the electron in
(6) this means that
$i^s=\stackrel{\ast}{\alpha^s}$, $k^s=
\stackrel{\ast}{\beta^s}$. The horizontal constraints connect the elements
$i$ and $k$. They correspond to Dirac's equations for free particles in
a momentum space [11].

{\bf 8. Momentum Space (Velocity Space)}

In the rank (3,3) BSCR, of element parameters -- two-component spinors --
the four-vectors {\it physically interpreted as velocity (or
momentum) components} of particles are conventionally constructed.
Introducing Dirac's four-row matrices in the corresponding representation,
the electron four-velocity components may be presented as follows:
\begin{eqnarray} u^0 & = & \frac{1}{2}(\overline{e}\gamma^0
e)=\frac{1}{2}(i^1\alpha^1+ i^2\alpha^2
+ k^1\beta^1+ k^2\beta^2); \nonumber \\
u^1 & = & \frac{1}{2}(\overline{e}\gamma^1 e)=
\frac{1}{2}(i^1\alpha^2+i^2\alpha^1+ k^1\beta^2+
k^2\beta^1); \\
u^2 & = & \frac{1}{2}(\overline{e}\gamma^3 e)=
\frac{i}{2}(i^1\alpha^2- i^2\alpha^1+ k^1\beta^2 -
k^2\beta^1); \nonumber \\
u^3 & = & \frac{1}{2}(\overline{e}\gamma^3 e)=
\frac{1}{2}(i^1\alpha^1-i^2\alpha^2+k^1\beta^1-
k^2\beta^2), \nonumber \end{eqnarray}
where $\overline{e}=e^{\dag}\gamma^0$. Defining the matrix $\gamma^5$, the
left and right components of the electron may be conventionally introduced,
then the left component of the electron is described by the pair of elements
$i$, $\alpha$, and the right component is described by the pair $k$, $\beta$.
It is evident that there is only one component for the neutrino. In the
general case, when the particles are described by a triple of elements, a
generalization of the above rules is necessary.

The above-mentioned transition from parameters to velocities may be
intrepreted as a transition from a binary geometry to a unary one by a
specific glueing of two pairs of elements of the two BSCR sets into new
elements of one set being described by Lobachevsky's geometry. It is easy to
verify that for one particle being characterized by the expression (6) we
have
\begin{equation} \left|\begin{array}{cc}
u_{i\alpha} & u_{i\beta} \\ u_{k\alpha} & u_{k\beta}
\end{array}\right|\equiv \left[{\alpha\atop i}{\beta\atop
k}\right]=g_{\mu\nu}u^{\mu}u^{\nu}. \end{equation}
For a free particle the four-dimensional velocities defined in such a way
possess the well-known property $ u^{\mu}u_{\mu}=Const=1.$

Introducing the second particle $e_2$ being described by the parameters: $j$,
$s$, $\gamma$, $\delta$, the scalar product of velocities (momenta) of two
particles may be defined as
\begin{equation} u^{\mu}_1u_{2\mu} = \frac{1}{4}
(\overline{e}_1\gamma^{\mu}e_1) (\overline{e}_2
\gamma_{\mu}e_2)=
\frac{1}{2}\left(\left[{\alpha\gamma\atop ij}\right]+
\left[{\alpha\delta\atop is}\right]+ \left[{\beta\gamma\atop
kj}\right]+ \left[{\beta\delta\atop ks}\right]\right),
\end{equation}
where the expression in square brackets denote the fundamental
$2\times 2$- rank (3,3) BSCR relations defined in (4).

For interacting particles the conditions of complex conjugation of elements
of the two sets are not satisfied any more. Introducing for particles in
initial and final states the complex conjugate quantities and constructing
from them, using the well-known formulas, four-velocities (momenta), we
arrive at the characteristics of particles before and after interaction
typical for the S-matrix.

{\bf 9. Choice of BSCR Rank}

In our works (see [11, 12]) properties of physical theories based on the
BSCR of ranks (2,2), (3,3), (4,4), (5,5), (6,6) and higher have been
analyzed step by step. The analysis has shown that the prototype of the known
kinds of physical interactions is constructed in the framework of {\it rank
(6,6) BSCR.} Therewith individual elementary particles (fermions) should be
described by triples of elements in each of two sets ${\cal M}$ and
${\cal N}$. This corresponds to the familiar notions of the structure of
baryons consisting of three quarks. In this approach these notions are
extended to leptons as well.

{\bf 10. Base $6\times 6$-Relations}

As a prototype of such related notions as the S-matrix or the action
(Lagrangian) for two interacting particles there should be some expression
containing two triples, i.e. six elements of one set ${\cal M}$ and six
elements of the set ${\cal N}$. Such is the {\it base $6\times 6$-relations}
being written in the framework of the rank (6,6) BSCR as follows
$$ \left\{{\alpha\atop i}{\beta\atop k}{\gamma\atop
j}{\delta\atop s}{\lambda\atop l}{\rho\atop r}\right\}\equiv
\left|\begin{array}{ccccccc}
0 & 1 & 1 & 1 & 1 & 1 & 1 \\
1 & u_{i\alpha} & u_{i\beta} & u_{i\gamma} & u_{i\delta} &
u_{i\lambda} & u_{i\rho} \\ 1 &
u_{k\alpha} & u_{k\beta} & u_{k\gamma} & u_{k\delta} &
u_{k\lambda} & u_{k\rho} \\ 1 & u_{j\alpha} & u_{j\beta} &
u_{j\gamma} & u_{j\delta} & u_{j\lambda} & u_{j\rho} \\ 1 &
u_{s\alpha} & u_{s\beta} & u_{s\gamma} & u_{s\delta} &
u_{s\lambda} & u_{s\rho} \\ 1 & u_{l\alpha} & u_{l\beta} &
u_{l\gamma} & u_{l\delta} & u_{l\lambda} & u_{l\rho} \\
1 & u_{r\alpha} & u_{r\beta} & u{r\gamma} & u_{r\delta} &
u_{r\lambda} & u_{r\rho} \end{array}\right|=$$
\begin{equation}=\left|\begin{array}{ccc|ccc}
i^1 & k^1 & j^1 & s^1 & l^1 & r^1 \\
i^2 & k^2 & j^2 & s^2 & l^2 & r^2 \\ \hline
i^3 & k^3 & j^3 & s^3 & l^3 & r^3 \\
i^4 & k^4 & j^4 & s^4 & l^4 & r^4 \\
i^5 & k^5 & j^5 & s^5 & l^5 & r^5 \\
1 & 1 & 1 & 1 & 1 & 1
\end{array}\right|\times
\left|\begin{array}{ccc|ccc}
\alpha^1 & \beta^1 & \gamma^1 & \delta^1 & \lambda^1 & \rho^1
\\
\alpha^2 & \beta^2 & \gamma^2 & \delta^2 & \lambda^2 & \rho^2
\\ \hline
\alpha^3 & \beta^3 & \gamma^3 & \delta^3 & \lambda^3 & \rho^3
\\
\alpha^4 & \beta^4 & \gamma^4 & \delta^4 & \lambda^4 & \rho^4
\\
\alpha^5 & \beta^5 & \gamma^5 & \delta^5 & \lambda^5 & \rho^5
\\ 1 & 1 & 1 & 1 & 1 & 1
\end{array}\right|, \end{equation}
where the vertical lines underline the fact that the first particle is
described by the elements $i$, $k$, $j$, $\alpha$, $\beta$, $\gamma$, and
the second one is described by the elements $s$, $l$, $r$, $\delta$,
$\lambda$, $\rho$. This expression is invariant under the transformations of
parameters of the group $SL(5,C)$ and is a specific volume in the binary
geometry of rank (6,6).

The physical interpretation of the base $6\times 6$-relation is illustrated
by the digrams of Fig. 2, where on the left the 12-tail structure of binary
geometrophysics is depicted. Two triples of the lower lines describe initial
states of two particles, and two upper triples -- their final states. A
generalization of Feynman's type diagram is presented in the middle, and
a standard diagram of particle scattering is given on the right.
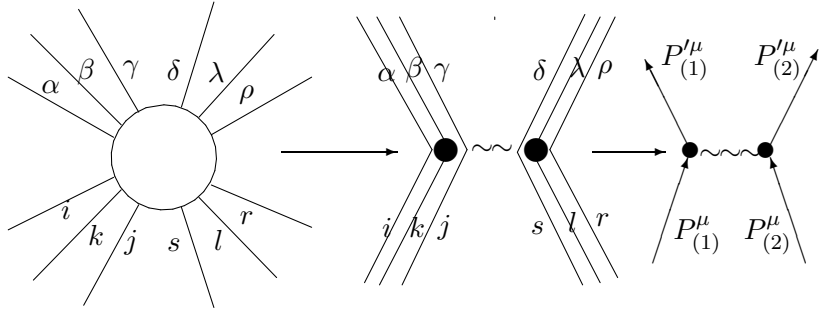
\begin{figure}[htbp]
\special{em:linewidth 0.4pt}
\unitlength 1mm
\linethickness{0.4pt}
\begin{center}
\begin{picture}(111.00,50.67)
\emline{24.00}{37.00}{1}{25.59}{36.82}{2}
\emline{25.59}{36.82}{3}{27.09}{36.28}{4}
\emline{27.09}{36.28}{5}{28.43}{35.42}{6}
\emline{28.43}{35.42}{7}{29.54}{34.27}{8}
\emline{29.54}{34.27}{9}{30.37}{32.91}{10}
\emline{30.37}{32.91}{11}{30.86}{31.39}{12}
\emline{30.86}{31.39}{13}{31.00}{29.80}{14}
\emline{31.00}{29.80}{15}{30.77}{28.22}{16}
\emline{30.77}{28.22}{17}{30.19}{26.73}{18}
\emline{30.19}{26.73}{19}{29.29}{25.42}{20}
\emline{29.29}{25.42}{21}{28.11}{24.34}{22}
\emline{28.11}{24.34}{23}{26.72}{23.55}{24}
\emline{26.72}{23.55}{25}{25.19}{23.10}{26}
\emline{25.19}{23.10}{27}{23.60}{23.01}{28}
\emline{23.60}{23.01}{29}{22.03}{23.28}{30}
\emline{22.03}{23.28}{31}{20.56}{23.90}{32}
\emline{20.56}{23.90}{33}{19.27}{24.84}{34}
\emline{19.27}{24.84}{35}{18.22}{26.05}{36}
\emline{18.22}{26.05}{37}{17.48}{27.46}{38}
\emline{17.48}{27.46}{39}{17.07}{29.00}{40}
\emline{17.07}{29.00}{41}{17.03}{30.60}{42}
\emline{17.03}{30.60}{43}{17.34}{32.16}{44}
\emline{17.34}{32.16}{45}{18.01}{33.61}{46}
\emline{18.01}{33.61}{47}{18.98}{34.88}{48}
\emline{18.98}{34.88}{49}{20.22}{35.89}{50}
\emline{20.22}{35.89}{51}{21.65}{36.59}{52}
\emline{21.65}{36.59}{53}{24.00}{37.00}{54}
\emline{18.33}{34.33}{55}{6.33}{46.33}{56}
\emline{20.67}{36.00}{57}{12.67}{49.67}{58}
\emline{26.33}{36.67}{59}{30.67}{50.67}{60}
\emline{28.67}{35.33}{61}{38.67}{46.33}{62}
\emline{30.33}{33.00}{63}{43.67}{40.67}{64}
\emline{18.33}{25.67}{65}{6.67}{13.67}{66}
\emline{20.67}{23.67}{67}{13.33}{10.33}{68}
\emline{26.33}{23.33}{69}{30.67}{10.00}{70}
\emline{28.67}{24.67}{71}{39.00}{14.00}{72}
\emline{30.33}{26.33}{73}{43.67}{20.67}{74}
\emline{52.67}{13.00}{75}{62.00}{31.00}{76}
\emline{62.00}{31.00}{77}{52.33}{48.67}{78}
\emline{55.67}{13.00}{79}{64.33}{31.00}{80}
\emline{64.33}{31.00}{81}{55.33}{48.67}{82}
\emline{79.67}{13.00}{83}{71.00}{30.67}{84}
\emline{71.00}{30.67}{85}{80.00}{49.00}{86}
\emline{82.00}{13.33}{87}{73.00}{30.67}{88}
\emline{73.00}{30.67}{89}{82.00}{48.67}{90}
\emline{84.33}{14.00}{91}{75.33}{31.00}{92}
\emline{75.33}{31.00}{93}{84.00}{48.67}{94}
\emline{68.00}{48.33}{95}{68.00}{48.67}{96}
\put(89.00,16.00){\vector(1,3){4.67}}
\put(94.00,31.00){\vector(-1,2){6.00}}
\put(109.33,16.00){\vector(-1,3){4.67}}
\put(104.33,31.00){\vector(1,2){6.67}}
\put(39.67,30.67){\vector(1,0){15.33}}
\put(81.00,30.67){\vector(1,0){9.67}}
\put(61.50,31.00){\circle*{3.00}}
\put(73.50,31.00){\circle*{3.00}}
\put(94.00,31.00){\circle*{2.00}}
\put(104.00,31.00){\circle*{2.00}}
\put(66.33,31.00){\makebox(0,0)[cc]{$\sim$}}
\put(69.00,31.00){\makebox(0,0)[cc]{$\sim$}}
\put(96.67,30.33){\makebox(0,0)[cc]{$\sim$}}
\put(99.33,30.33){\makebox(0,0)[cc]{$\sim$}}
\put(102.00,30.33){\makebox(0,0)[cc]{$\sim$}}
\put(13.67,41.33){\makebox(0,0)[cc]{$\beta$}}
\put(19.67,41.67){\makebox(0,0)[cc]{$\gamma$}}
\put(25.33,41.67){\makebox(0,0)[cc]{$\delta$}}
\put(31.00,41.33){\makebox(0,0)[cc]{$\lambda$}}
\put(35.00,38.33){\makebox(0,0)[cc]{$\rho$}}
\put(11.00,23.00){\makebox(0,0)[cc]{$i$}}
\put(25.33,18.33){\makebox(0,0)[cc]{$s$}}
\put(31.33,19.00){\makebox(0,0)[cc]{$l$}}
\put(35.00,22.00){\makebox(0,0)[cc]{$r$}}
\put(19.67,18.67){\makebox(0,0)[cc]{$j$}}
\put(53.67,41.00){\makebox(0,0)[cc]{$\alpha$}}
\put(61.00,41.33){\makebox(0,0)[cc]{$\gamma$}}
\put(74.00,41.67){\makebox(0,0)[cc]{$\delta$}}
\put(79.00,42.00){\makebox(0,0)[cc]{$\lambda$}}
\put(82.67,41.67){\makebox(0,0)[cc]{$\rho$}}
\put(53.67,20.67){\makebox(0,0)[cc]{$i$}}
\put(61.33,20.67){\makebox(0,0)[cc]{$j$}}
\put(73.67,20.67){\makebox(0,0)[cc]{$s$}}
\put(78.33,21.33){\makebox(0,0)[cc]{$l$}}
\put(82.33,21.67){\makebox(0,0)[cc]{$r$}}
\put(95.00,19.67){\makebox(0,0)[cc]{$P^{\mu}_{(1)}$}}
\put(93.33,43.00){\makebox(0,0)[cc]{$P^{\prime\mu}_{(1)}$}}
\put(104.33,19.67){\makebox(0,0)[cc]{$P^{\mu}_{(2)}$}}
\put(105.67,43.00){\makebox(0,0)[cc]{$P^{\prime\mu}_{(2)}$}}
\emline{17.33}{32.67}{97}{3.33}{40.67}{98}
\emline{17.33}{27.33}{99}{3.33}{20.33}{100}
\emline{50.67}{13.33}{101}{59.67}{31.00}{102}
\emline{59.67}{31.00}{103}{49.67}{48.67}{104}
\put(9.00,39.33){\makebox(0,0)[cc]{$\alpha$}}
\put(15.00,19.67){\makebox(0,0)[cc]{$k$}}
\put(57.67,20.67){\makebox(0,0)[cc]{$k$}}
\put(57.33,41.67){\makebox(0,0)[cc]{$\beta$}}
\end{picture}
\end{center}
\vspace{-10mm}
\caption{Physical illustration of the base $6\times 6$-relation.}\end{figure}

\section{Transition from the Bionary Volume to the S-Matrix Prototype or the
Action (Lagrangian)}
\hspace*{\parindent}
To pass from the base $6\times 6$-relation to prototypes of the S-matrix or
the action (lagrangian) of two particles, the following set of approaches,
procedures and principles should be used.

{\bf 1. Splitting Procedures}

First of all, it is necessary to perform a procedure of splitting (or
reduction) corresponding to the $4+s$-splitting procedure into
four-dimensional spacetime and additional dimensions in multidimensional
geometrical models of Kaluza-Klein's type theory. It lies in separating the
parameters with indices 1 and 2 called {\it external ones}, from three rest
parameters with indices 3, 4, 5 called {\it internal ones}. From the external
parameters four-dimensional momenta (velocities) are constructed both in the
framework of the rank (3,3) BSCR according (7), and from the internal ones
charges of elementary particles are constructed. In fact, this procedure
means splitting the initial rank (6,6) BSCR into two subsystems: a rank (3,3)
BSCR (with two parameters) and a rank (4,4) BSCR (with three parameters). As
a result of splitting, the initial transformation group $SL(5,C)$ is narrowed
down up to two subgroups: $SL(2,C)$ -- for external parameters and $SL(3,C)$
(or a narrower one $SU(3)$) -- for internal parameters.

To construct a prototype of the action, the base $6\times 6$-relation should
be represented in the form reduced to four-dimensional spacetime when
parameters with indices 1 and 2 are singled out, and the final expression has
the Lorentz-invariant ($SL(2,C)$-invariant) form. This may be performed by
expansion of the determinants on the right in (10) with repect first two
lines. Multiplying them, we arrive at a set of 225 expressions of the form
\begin{equation}
\left\{{\alpha\atop i}{\beta\atop k}{\gamma\atop
j}{\delta\atop s}{\lambda\atop l}{\rho\atop r}\right\}
=\sum^{225}\left[{\alpha\atop i}{\delta\atop s}\right]
\left({\beta\atop k}{\gamma\atop j}{\lambda\atop l}
{\rho\atop r}\right), \end{equation}
where the square brackets denote the fundamental $2\times 2$-relations
constructed from parameters with indices 1 and 2, and the paretheses denote
combinations of internal parameters of the form
\begin{equation} \left({\beta\atop k}{\gamma\atop j}
{\lambda\atop l}{\rho\atop r}\right)=\left|\begin{array}{cc|cc}
k^3 & j^3 & l^3 & r^3 \\ k^4 & j^4 & l^4 & r^4 \\ k^5 & j^5 &
l^5 & r^5 \\ 1 & 1 & 1 & 1 \end{array}\right|\times
\left|\begin{array}{cc|cc} \beta^3 & \gamma^3 & \lambda^3 &
\rho^3 \\ \beta^4 & \gamma^4 & \lambda^4 & \rho^4 \\ \beta^5 &
\gamma^5 & \lambda^5 & \rho^5 \\ 1 & 1 & 1 & 1
\end{array}\right|. \end{equation}

{\bf 2. Classification and Physical Interpretation of Terms of the Base
$6\times 6$-Relation}

The set of 225 terms of the form (11) should be divided into 9 subsets
depicted as subblocks of a $15\times 15$-matrix separated by horizontal and
vertical lines:
$$\left\{{\alpha\atop i}{\beta\atop k}{\gamma\atop
j}{\delta\atop s}{\lambda\atop l}{\rho\atop r}\right\}=
\left(\begin{array}{ccc|ccccccccc|ccc}
\star & \star & \star & \cdot & \cdot & \cdot & \cdot & \cdot &
\cdot & \cdot & \cdot & \cdot & \star & \star & \star \\
\star & \star & \star & \cdot & \cdot & \cdot & \cdot & \cdot &
\cdot & \cdot & \cdot & \cdot & \star & \star & \star \\
\star & \star & \star & \cdot & \cdot & \cdot & \cdot & \cdot &
\cdot & \cdot & \cdot & \cdot & \star & \star & \star \\ \hline
\cdot & \cdot & \cdot & \star & \cdot & \cdot & \cdot & \cdot &
\cdot & \cdot & \cdot & \cdot & \cdot & \cdot & \cdot \\
\cdot & \cdot & \cdot & \cdot & \star & \cdot & \star & \cdot &
\cdot & \cdot & \cdot & \cdot & \cdot & \cdot & \cdot \\
\cdot & \cdot & \cdot & \cdot & \cdot & \star & \cdot & \cdot &
\cdot & \star & \cdot & \cdot & \cdot & \cdot & \cdot \\
\cdot & \cdot & \cdot & \cdot & \star & \cdot & \star & \cdot &
\cdot & \cdot & \cdot & \cdot & \cdot & \cdot & \cdot \\
\cdot & \cdot & \cdot & \cdot & \cdot & \cdot & \cdot & \star &
\cdot & \cdot & \cdot & \cdot & \cdot & \cdot & \cdot \\
\cdot & \cdot & \cdot & \cdot & \cdot & \cdot & \cdot & \cdot &
\star & \cdot & \star & \cdot & \cdot & \cdot & \cdot \\
\cdot & \cdot & \cdot & \cdot & \cdot & \star & \cdot & \cdot &
\cdot & \star & \cdot & \cdot & \cdot & \cdot & \cdot \\
\cdot & \cdot & \cdot & \cdot & \cdot & \cdot & \cdot & \cdot &
\star & \cdot & \star & \cdot & \cdot & \cdot & \cdot \\
\cdot & \cdot & \cdot & \cdot & \cdot & \cdot & \cdot & \cdot &
\cdot & \cdot & \cdot & \star & \cdot & \cdot & \cdot \\ \hline
\star & \star & \star & \cdot & \cdot & \cdot & \cdot & \cdot &
\cdot & \cdot & \cdot & \cdot & \star & \star & \star \\
\star & \star & \star & \cdot & \cdot & \cdot & \cdot & \cdot &
\cdot & \cdot & \cdot & \cdot & \star & \star & \star \\
\star & \star & \star & \cdot & \cdot & \cdot & \cdot & \cdot &
\cdot & \cdot & \cdot & \cdot & \star & \star & \star
\end{array}\right)\equiv $$
\begin{equation}\equiv \left(\begin{array}{c|c|c}
M(4,0) & +M(3,1) & +M[(2)(2)]+ \\ \\ \hline
+M(3,1) & +M(2,2) & +M(1,3)+ \\ \\ \hline
+M[(2),(2)] & +M(1,3) & +M(0,4) \end{array}\right),
\end{equation}
where individual terms are denoted by points or astersks.
The latter mark terms of utmost importance in this theory (with a special
order of terms).

The subblocks differ in the character of the fundamental
$2\times 2$-relations. The middle $9\times 9$-subblock contains terms of the
form $\left[{\cdot\atop\cdot} {|\atop |}{\cdot\atop\cdot}\right]$, where the
vertical line divided the parameters characterizing two particles. Such terms
describe vector-vector interactions of two pareticles. In the gauge theory
they  correspond to the interactions via intermediate vector bosons
(gluons, photons, Z- or W-bosons).

The left lower or right upper $3\times 3$-subblocks $M[(2),(2)]$ contain
terms of the form $\left[{\cdot \ \cdot\atop\overline{\cdot \ \cdot}}
\right]$, where the horizontal line divides the parameters characterizing
two particles. Such terms describe scalar interactions of two particles. They
correspond to Higgs' scalar bosons. These subblocks will be called mass ones.

The diagonal left upper $M(4,0)$ and right lower $M(0,4)$ subblocks contain
external parameters of only one of the particles. They correspond to the
action (Lagrangian) of ``free'' particles. Such submartices will be called
free for the first and second particles.

The rest four extreme $3\times 9$- и $9\times 3$-subblocks contain three
parameters of one particle and one parameter of another particle. There are
criteria allowing contributions of such terms to be excluded.

{\bf 3. Unified Principle of Describing Baryons and Leptons}

The theory being presented here permits interactions (processes involving)
baryons, massive leptons and neutrinos to be described in the same way.
All particles mentioned are described by triples of elements in each of the
two sets (states). The baryons are generally characterized by all three
two-component columns of the external parameters containing nonzero
components. Thus, e.g., in the set ${\cal M}$ the baryon (b) is characterized
by a rectangular $3\times 5$-matrix of the parameters of its consistuents:
\begin{equation} (b)\Rightarrow\left(\begin{array}{ccc}
i^1 & k^1 & j^1 \\ i^2 & k^2 & j^2  \\ \hline  i^3 & k^3 & j^3
\\ i^4 & k^4 & j^4 \\ i^5 & k^5 & j^5 \end{array}\right)\equiv
\left(\begin{array}{ccc} i^1 & k^1 & j^1 \\ i^2 & k^2 & j^2 \\
\hline  c^3_{(1)} & c^3_{(2)} & c^3_{(3)} \\
c^4_{(1)} & c^4_{(2)} & c^4_{(3)} \\
c^5_{(1)} & c^5_{(2)} & c^5_{(3)} \end{array}\right),
\end{equation}
where two upper lines correspond to external two-component spinors, and the
three rest -- to additional parameters (three-component Finsler spinors)
describing internal degrees of freedom (charges) of an interacting particle.

For massive leptons (electrons (e)) the matrix has the same form (14),
however, one of the upper two-component columns consists of zero external
parameters. Let it be the third column: $j^1=j^2=0$. For neutrinos
$(\nu)$, being also described by $\times 5$-матрицей (14), two two-component
columns of the external parameters are zero. Let it be the first two columns:
$i^1=i^2=0$, $k^1=k^2=0$. For the leptons defined in such a way the number
nonzero terms in the base $6\times 6$-relation (13) reduces considerably. It
should be noted that the number zero columns of the external parameters is an
invariant property of particles with respect to distinguished parameter
transfomation groups. The particle definitions by external parameters given
here are in agreement with those presented in paragraph 7 of the previous
section.

{\bf 4. Exchange Character of Physical Interactions}

If one write the base $6\times 6$-relation for two particles with the same
internal parameters, then it vanishes due to antisymmetry of determinant
columns. A nonzero result arises in using the exchange mechanism of physical
interactions. It is based on the postulate that, according to values of the
external parameters, the particles may be in two kinds of states: in the
$U$-state (``normal'') or in one of the $X$-states (``excited''). The
interaction process consists in an exchange of states between particles.

The ``normal'' or $U$-state is characterized by a nonzero
$3\times 3$-determinant of the internal parameters. The analysis of possible
simplest kinds of $X$-states (taking account of the principle of
correspondence to the conventional theory) shows that the $X$-states may be
of two kinds. One type ($X_C$) is also characterized by a nonzero determinant
of three columns of the external parameters, and for another type ($X_N$)
such a deteminant is zero.

{\bf 4a) $X_C$-states (charged)}

The simplest variant of determining $X_C$ states leading to nonzero base
$6\times 6$-relations lies in changing the sign of of the three columns of
the internal parameters of the $U$-state. In all there are three such
possibilities (channels) which, according to the column number with changed
sign, are called $X_X$-, $X_Y$- and $X_Z$-states. It can be shown that for
these $X_C$-states only three pairs of terms in the $M(2,2)$-matrix prove to
be nonzero. They are located in its $3\times 3$-subblocks similarly to
nonzero elements in six nondiagonal Gell-Mann's matrices $\lambda_n$. Such
terms correspond to interactions via charged vector bosons in the
conventional theory.

{\bf 4b) $X_N$-states (neutral)}

In $X_N$-states a pair or all three columns (three-dimensional vectors) of
the internal parameters are collinear. In the simplest case on may assume
that all three vectors (of the column) $\vec{c}^{\prime}_{(s)}$, where
$s=1,2,3$, are collinear, i.e. representable as
$ \vec{c}^{\prime}_{(s)}=C^{\prime}_s\vec{c}^{\prime}$,
where $\vec{c}^{\prime}$ is a three vector, and $C^{\prime}_s$ are three
coefficients. The coefficients $C^{\prime}_s$ in diagonal summands of the
submatrix $M(2,2)$ may be shown to enter only as differences, i.e. only two
combinations of them are independent. One may choose the combinations as
follows:
\begin{equation} C^{\prime}=\frac{1}{2}(C^{\prime}_1+
C^{\prime}_2-2C^{\prime}_3); \ \ \ \tilde{C}^{\prime}= -
\frac{1}{2}(C^{\prime}_1-C^{\prime}_2). \end{equation}
Two particular cases: А) $C^{\prime}\neq 0; \ \
\tilde{C}^{\prime}=0$; и В) $C^{\prime}= 0; \ \
\tilde{C}^{\prime} \neq 0$ define two channels: an А-channel with the
corresponding $X_A$-state and  В) a channel with the corresponding
$X_B$-state, which should be compared to two channels of interactions via
neutral vector bosons in the conventional theory. The two combinations of the
coefficients in (15) correspond to two Gell-Mann's diagonal matrices:
$\lambda_3$ and $\lambda_8$ in the conventional representation.

The above considerations characterize, so far only qualitatively, an essence
of five channels (three charged and two neutral) in the known types of
physical interactions. Details will be presented in the next article.

{\bf 5. Intermediate Vector Boson Treatment}

In the given approach, corresponding to (Fokker-Planck's)
action-at-a-distance concept, there are not intermediate carriers of
interactions (vector bosons). To them correspond the above channels of
interactions (types of $X$-states of elementary particles). For strong
interactions eight gluons correspond to these channels: three pairs of
charged gluons corresponding to the $X_X$-, $X_Y$- and $X_Z$-states, and two
neutral А- и В-gluons corresponding to the states $X_A$ and $X_B$. Similarly
``intermediate'' vector bosons ``carrying'' electroweak interactions.

{\bf 6. ``Matreshka'' Principle}

In the conventional gauge field theory three types of spaces are, in fact,
used to describe interactions: 1) an external one corresponding to the
classical four-dimensional spacetime  with the Lorentz transformation group
($SL(2,C)$ group), 2) an internal space of electroweak interactions in which
there takes place the group $SU(2)\times U(1)$ and 3) an inernal (chromatic)
space of stong interactions with the group $SU(3)$. The theory is being
constructed as a composition of these spaces, which is called
{\it the ``bricks'' principle}

Binary geometrophysics assumes a more economical approach of electroweak
interactions being considered as truncated strong interactions, which is
achieved by fixing one of the lines of the internal parameters in (14)
(let it will be a line of the parameters with index 3), when the group
$SL(3,C)$ (or $SU(3)$) of admissible transformations of the internal
parameters is narrowed down up to the subgroup $SL(2,C)$ (or $SU(2)$) of the
internal parameters with indices 4 and 5. It turns out that in such a
truncated theory there remain valid considerations presented above of two
types of interactions -- analogues of those via charged and neutral vector
bosons, with the difference that for one particle generation only one
channel (of three $X_C$) survives, which corresponds to interactions via
one pair of charged vector bosons.

In such a theory, electromagnetic interactions may be shown to correspond to
strong ones via neutral A-gluons, weak interactions via neutral vector
Z-bosons to strong ones via B-gluons, and electroweak interactions via
charged vector $W^{\pm}$-bosons to stong ones via one of the pairs of
charged vector gluons (say, $X^{\pm}$-gluons).

Such a principle of actual embedding of one type of physical interation
to another may be called the {\it ``matreshka'' princile}, an alternative to
the conventional ``bricks'' principle.

{\bf 7. Physical Interaction Channel Symmetry Principle}

The above procedures and principles allow prototypes of the action
(Lagrangian) of strong and electroweak interaction of baryons in terms of
quarks (their consistuents) as well as electroweak interaction of leptons
(massive ones and neutrinos) between one another and with baryons (quarks) to
be written down in the same way. These prototypes are constructed as
combinations of products of the four-velocities (of the external parameters)
of the particle (quark) components with the corresponding coefficients of the
internal parameters having a physical meaning of charges in strong and
electroweak interactions. However, therewith there remain uncertain
quantities and values of the independent constants and charges in the
corresponding interactions. This gap is eliminated by using the physical
interaction channel symmetry.

It turns out that for obtaining the familiar relations between charges it
suffices to know, first, the fact of an existence of the above channels,
second, the character of the presence of charges in the corresponding
interactions (multiplicativity in charges in the interactions via charged
vector bosons) and, third, the conditions of total symmetry of the above
channels for all quarks in strong interactions or for left components of
particles (quarks or leptons) in electroweak interactions. Therewith strong
interactions are unambigously shown to be characterized by only one constant
corresponding to the constant $g_0$ in chromodynamics, and electroweak ones
-- by two constants corresponding to the familiar charges $g_1$ and $g_2$ in
Weinberg-Salam's model.

In so doing it is natural that one relationship for two interaction
constants is written down, which permits Weiberg's angle to be calculated
theoretically.

{\bf 8. Description of Elementary Particle Generations}

The approach proposed (in the framework of binary geometrophysics) allows
one to resolve (or opens a new avenue of attack on) a number of problems of
the modern theory of physical interactions. For example, there opens a
possibility of theoretical justification of the existence of just three
generations of elementary particles. Simpifying the situation, one may argue
that the availabitity of three generations of elementary particles (quarks
and leptons) is closely related to the existence of three channels of strong
interactions of quarks via charged vector bosons. As mentioned here, in
electroweak interactions of the particles of the same generation an analogue
of only one pair of charged gluons in the form of W-bosons; in interactions
of the particles of two other generations there appear analogues of two other
pairs of charged gluons. For leptons, in particular, the definitions of three
generations are also related to arrangement of zero columns of the external
parameters.

It should be particularly emphasized that the approach proposed corresponds
basically to the conventional gauge models of physical interactions (e.g.,
see [22, 23]). In its framework notions of antiquarks and antileptons are
easily defined, besides introduced are mesons as well as particles being
described by pairs of elements in each of two sets of the rank (6,6) BSCR,
with, to fit the conventional approach, on of these elements corresponding to
the definition of a quark, and another -- of an antiquark. A number of other
problems will be elucidated in a succession of articles.

\section{Architectonics of Binary Geometrophysics}
\hspace*{\parindent}
The aforesaid relates only to the fundamentals of binary geometrophysics. The
simplest act of evolution -- a transition of two interacting particles
from one state to another has been discussed, with elementary particles being
as elementary bases for the latter. This always meant three specific subsets
of elements comprising the first and second interacting particles and
elementary basis.

An importamt problem of binary geometrophysics program lies in constructing
a transition from an algebraic theory of the level $R_{\mu}(\mu)$ to the
conventional spatial-temporal and other attendant notions of modern physics.
The most essential points here are as follows.

First of all, to the three specific susets of elements mentioned the fourth
subset, formed by all other elements that did not enter in these three
subsets, should be added. The first three of these subsets may be simple
(individual particles) or rather complex up to macro-objects, whereas the
added fourth subset is extremely complex by definition.

To turn to the conventional physical theories, it is necessary to take the
steps describable as the following chain:
\begin{equation} R_{\mu}(\mu)\stackrel{1}{\rightarrow}
R^{\prime}_{\mu}(\mu)\stackrel{2}{\rightarrow}
R^M_{\mu}(\mu)\stackrel{3}{\rightarrow}
R^M_m(\mu)\stackrel{4}{\rightarrow} R^M_m(m).
\end{equation}
This route has been described in our book [12] on the basis of a simplified
model in the framework of the rank (4,4) BSCR. In such an approach the base
$6\times 6$-relations (or $4\times 4$-relations in the simplified model)
generate an interval squared of Kaluza-Klein's multidimensional metric:
$$ (\mbox{Base} \ 6\times 6-\mbox{relation})\rightarrow
d\Sigma^2=G_{AB}dx^Adx^B, $$
where $G_{AB}$ are the multidimensional metric tensor components, and indices
$A$ and $B$ run the values: 0, 1, 2, 3, 4, 5, $\cdots$. Therewith part of
terms with the indices $A=\mu$ and $B=\nu$, taking the values: 0, 1, 2, 3,
is due to an angular diagonal subblock of the base $6\times 6$-relation
$M(4,0)$, the terms with mixed indices (one index is four-dimensional, and
another index with the values 5, 6, $\cdots$) are due to the middle submatrix
$M(2,2)$. An especial role plays diagonal components of the multidimensional
metric with indices $A,B>3$.

The prototypes of coordinate shifts are formally introduced in terms of
four-dimensional momenta $p^{\mu}=mdx^{\mu}/d\Sigma $, where
$d\Sigma=dx^4$ is the prototype of the multidimensional interval. The
square-law characteristic of the multidimensional metric prototype is due to
the square-law of particle momenta in a number of terms of the base
$6\times 6$-relations. The additional momentum components and shift
prototypes are obtained from the internal parameters of elements.

{\bf 1. The first step}, in fact, has been discussed in the previous section
of the article. It consisted in passing from the general notions of the theory
of relations between abstract elements to prototypes of the action
(Lagrangian) of interacting particles and other conventional physical
notions. This step may be denoted as a transition
$R_{\mu}(\mu)\stackrel{1}{\rightarrow} R^{\prime}_{\mu}(\mu)$,
where $R^{\prime}_{\mu}(\mu)$ means the theory obtained from rank (6,6) BSCR,
as a result of applying the procedures and principles of the previous section.

At this stage being described in algebraic terms there do not exist most of
the habitual notions of modern physics. First of all, there is no a notion of
the classical spacetime as well as many attendant macronotions: there are no
metric, causality, world lines; the notions of wave functions, particles and
fields -- interaction carriers are absent; there are no field propagators and
singular functions troubling physicists very much in the twentieth century.

There are no many other notions, which makes the conventional field theory
unthinkable. Finally, gravitational interactions are absent at this level.

{\bf 2.  The second step (chain)} consists in taking account of all particle
of the surrounding world. In modern literature this is called Mach's principle
[24]. This step may denoted as a transition
$ R^{\prime}_{\mu}(\mu)\stackrel{2}{\rightarrow}
R^M_{\mu}(\mu),$ where to the two above factors we added one more, i.e. the
external world, designated by the subscript $M$. The basic idea of this step
lies in a transition from one base $r\times r$-relation to the sum of such
base relations which necessarily contain a fixed number each of elements of
an isolated (first) particle in each of two BSCR sets ${\cal M}$ и ${\cal N}$.
Symbolically, we represent this transition by the formula:
\begin{equation} \left\{{\alpha\atop i}{\beta\atop
k}{\gamma\atop j}{|\atop |}{\delta\atop s}{\lambda\atop
l}{\rho\atop r}\right\}^{\prime}\rightarrow
\Sigma(1,1^{\prime})=\sum^{World}_{s,l,r;
\delta\lambda\rho}\left\{{\alpha\atop i}{\beta\atop
k}{\gamma\atop j}{|\atop |}{\delta\atop s}{\lambda\atop
l}{\rho\atop r}\right\}^{\prime}, \end{equation}
where the prime means the base $6\times 6$-relations interpreted according to
the previous section. Here the summing is performed over all elements of two
BSCR sets ${\cal M}$ and ${\cal N}$, except for elements determining the
first particle, i.e. over all particles of the surrounding world.

In this case we are dealing with a set of second particles (but not one
particle) each of which contributes to the value of the prototype of
multidimensional interval of the first particle. It may be singled out by
considering separately contributions of the base $6\times 6$-relations
containing elements of the second particle.

In the framework of this step (stage of theory development) one needs to
perform the procedure of $4+1+\cdots$-splitting corresponding to reducibility
(splitting) of the rank (6,6) BSCR into two subsystems of ranks (3,3) and
(4.4). In Kaluza-Klein's multidimensional theory context this procedure
corresponds to employment of the monadic, dyadic, $\cdots$, s-аdic methods of
splitting [14, 18] of the n-dimensional manifold into the four-dimensional
spacetime and additional dimensions orthogonal to it.

It should be especially emphasized that {\bf only after application of the
splitting procedure one may speak about appearance of gravitational
interaction.} In the expression (16) the contributions of other particles are
due to electroweak interactions. After the splitting procedure there arise
prototypes of the components of the four-dimensional curved metric being
defined by a sum of Minkowski space  and quadratic contributions of mixed
components of the metric $G_{\mu a}$, where $a=4, 5, \cdots$, which
corresponds to the form of Kaluza-Klein's metric [14]. Therewith the
resulting components of the metric for additional dimensions, physically
interpretable as prototypes of the electromagnetic and other ``intermediate
vector fields'' field strengths, contain sums of linear contributions of
surrounding particles.

Thus, {\it both electromagnetic and gravitational interactions are determined
by the same contributions, but they are summing up linearly for an
electromagnetic interaction, and quadratically for a gravitational one}.
Remind that the square of a sum of terms is inequal to a sum of squares of
the same terms. This means that a linear sum of contributions of
electromagnetic interaction may be zero, whereas the sum of squares of the
same contributions will be always nonzero, i.e. gravitation interaction will
exist even in the absence of an observable electromagnetic field. This
conclusion determines a new glance to the nature of gravity as well as the
essence of Einstein's General Relativity. In particular, this determines
another approach to many fundamental problem of the modern theory of gravity.

After the first stage and performing the splitting procedure in the theory
there arise large sums of the terms corresponding to the surrounding world.
In Wheeler-Feynman's absorber theory [16], in the framework of
Fokker-Planck's at-a-distance theory, such sums are, in fact,
approximated by integrating over matter uniformly distributed in the whole
space.

{\bf 3. The third step} consists in a transition from the elementary bases
$\mu$ to rather complex systems of base elements and in the limit to a
macroinstrument $m$: $R^M_{\mu}(\mu)\stackrel{3}{\rightarrow} R^M_m(\mu),$
which, in fact, corresponds to quantum mechanics. Symbollically, this step
may be described by the formula:
\begin{equation} \Sigma(1,1^{\prime})\rightarrow
\sum_{basis}\left(\sum^{World}_{s,l,r;
\delta\lambda\rho}\left\{{\alpha\atop i}{\beta\atop
k}{\gamma\atop j}{|\atop |}{\delta\atop s}{\lambda\atop
l}{\rho\atop r}\right\}^{\prime}\right)\rightarrow
d\Sigma^2(1,1^{\prime}). \end{equation}
Here there arise one more summing over the elementary bases comprising a
macroinstrument. The sums are complicated considerably, however, at this
stage the new sums are approximated by integrating over the momenta of
transfer from some objects to others. These sums correspond to Fourier
integrals in the conventional representations of vector field potentials.

Only after performing the second stage there arise a possibility of speaking
about the wave functions of spinor particles and about boson fields --
interaction carriers. At this stage there arise notions of retarded and
advanced interactions (potentials). The latter are eliminated according to
Feynman-Wheeler's procedure of allowing for world's absolute absorber. At
this stage there arise quantum mechanics and quantum field theory.

{\bf 4. The fourth step} consists in a transition from consideration of
microparticles to description of micro-objects:
$R^M_m(\mu)\stackrel{4}{\rightarrow} R^M_m(m),$ which means a transition to
classical physics. The latter is realized by summing (averaginig) over all
particles comprising macro-objects (a isolated particle). Strictly speaking,
only after this stage the classical spacetime is said to be constructed from
primary relations. It is here that there arise notions of distances and time
intervals meant in the previous paragraph, when speaking of the wave
functions and boson interaction field-carriers.

\section{Conclusion}
\hspace*{\parindent}
In the present article the main principles forming the basis for a new
approach to constructing the unified theory of spacetime and physical
interactions have been indicated, only a brief characteristic of the most
essential results obtained in the framework of binary geometrophysics are
given. A thorough presentation will be given in a run of articles devoted
to separate above-mentioned problems.

The author thanks the participants of the Russian Gravitational Society
Seminar at Physics Faculty, Moscow State University, as well as the
participants of the Theoretical Physics Laboratory Seminar at Joint Institute
for Nuclear Research in Dubna for detailed discussions and valuable comments.

\end{document}